\documentclass[letterpaper, 10 pt, conference]{ieeeconf} 

\IEEEoverridecommandlockouts

\usepackage{textcomp}
\usepackage{amsmath} 
\usepackage{amssymb} 
\usepackage{graphicx}
\usepackage{url}
\usepackage{lipsum}
\usepackage{tabularx}
\usepackage{booktabs}
\usepackage[caption=false,font=footnotesize]{subfig}
\usepackage{xparse}% http://ctan.org/pkg/xparse
\NewDocumentCommand{\rot}{O{90} O{1em} m}{\makebox[#2][l]{\rotatebox{#1}{#3}}}%
\usepackage{multirow}
\usepackage{tikz}

\newcommand\copyrighttext{%
  \footnotesize \textcopyright 2020 IEEE.  Personal use of this material
  is permitted. Permission from IEEE must be obtained for all other uses, in
  any current or future media, including reprinting/republishing this material
  for advertising or promotional purposes, creating new collective works, for
  resale or redistribution to servers or lists, or reuse of any copyrighted
  component of this work in other works}
\newcommand\copyrightnotice{%
  \begin{tikzpicture}[remember picture,overlay]
    \node[anchor=south,yshift=10pt] at (current page.south) {\fbox{\parbox{\dimexpr\textwidth-\fboxsep-\fboxrule\relax}{\copyrighttext}}};
  \end{tikzpicture}%
  }

\newcommand{\x}{{\mathbf{x}}}

\title{\bf \LARGE Inferring the Spatial Distribution of Physical
  Activity in Children Population from Characteristics of the
  Environment}

\author{Ioannis Sarafis, Christos Diou, Vasileios Papapanagiotou, 
  Leonidas Alagialoglou and Anastasios
  Delopoulos\\ \thanks{ *The work leading to these results
    has received funding from the European Community's Health,
    demographic change and well-being Programme under Grant Agreement
    No. 727688 \protect\url{(https://bigoprogram.eu)}, 01/12/2016 -
    30/11/2020.  The Institution's Ethical Review Board approved all
    experimental procedures involving human subjects.}%
  \thanks{All authors are with Multimedia Understanding Group, Information Processing
    Laboratory, Aristotle University of Thessaloniki, Greece.}}

\begin{document}

\maketitle
\thispagestyle{empty}
\pagestyle{empty}
\copyrightnotice

%%%%%%%%%%%%%%%%%%%%%%%%%%%%%%%%%%%%%%%%%%%%%%%%%%%%%%%%%%%%%%%%%%%%%%%%%%%%%%%%
\begin{abstract}
Obesity affects a rising percentage of the children and adolescent
population, contributing to decreased quality of life and increased
risk for comorbidities. Although the major causes of obesity are
known, the obesogenic behaviors manifest as a result of complex
interactions of the individual with the living environment. For this
reason, addressing childhood obesity remains a challenging problem for
public health authorities. The BigO project (https://bigoprogram.eu)
relies on large-scale behavioral and environmental data collection to
create tools that support policy making and intervention design. In
this work, we propose a novel analysis approach for modeling the
expected population behavior as a function of the local
environment. We experimentally evaluate this approach in predicting
the expected physical activity level in small geographic regions using
urban environment characteristics. Experiments on data collected from
156 children and adolescents verify the potential of the proposed
approach. Specifically, we train models that predict the physical
activity level in a region, achieving 81\% leave-one-out accuracy. In
addition, we exploit the model predictions to automatically visualize
heatmaps of the expected population behavior in areas of interest,
from which we draw useful insights. Overall, the predictive models and
the automatic heatmaps are promising tools in gaining direct
perception for the spatial distribution of the population's behavior,
with potential uses by public health authorities.
\end{abstract}

%%%%%%%%%%%%%%%%%%%%%%%%%%%%%%%%%%%%%%%%%%%%%%%%%%%%%%%%%%%%%%%%%%%%%%%%%%%%%%%%
\section{INTRODUCTION}
\label{sec:introduction}

The obesity epidemic soared in the recent decades, affecting a
significant percentage of the global population. Obesity is increasing
rapidly in the younger population, with the World Health Organization
(WHO) estimating that over $18\%$ of the children and adolescent
population aged $5$-$19$ were overweight or obese in $2016$, compared
to $4\%$ in $1975$ \cite{who2016obesity}.

At the individual's level, the major cause of obesity is the imbalance
between energy expenditure and energy intake
\cite{butte2007energy}. However, behind this imbalance, there are
complex behavior patterns and interactions associated with the living
environment \cite{romieu2017energy_short}. To this end, public health
authorities are tasked with designing policies that create ``active
environments'' and promote healthier lifestyle for the population
\cite{world2019global, gilescorti2016city_short}.

Towards this goal, the \emph{BigO: Big Data Against Childhood
  Obesity}$^*$ project takes advantage of new data sources to create
novel tools for policy making and intervention design
\cite{diou2019bigo_short,sarafis2019behaviour}. BigO relies on: 1)
large-scale, objective monitoring of the population's behavior, in
real-life conditions, using commodity devices (e.g., smartphones and
smartwatches); 2) estimating the environment characteristics using
open and online data sources; and 3) associating population behaviors
with environment characteristics to create actionable knowledge for
public health authorities. An important aspect of the BigO
methodologies is that the data are anonymized and aggregated in
geographical regions in order to increase privacy protection for
the participating individuals.

In this work, we present an analysis approach developed in the context
of BigO for predicting behaviors in children and adolescent population
using measurements of the environment. Then, we demonstrate the
proposed approach in building models for predicting the expected
physical activity level. The predictive models use measurements of the
urban environment in relatively small geographic regions
(approx. $0.57$ km$^2$) and predict the expected physical activity
level for the resident population, as High or Low level.

Experiments are conducted on data collected from $156$ children and
adolescents during BigO pilots in the Thessaloniki metropolitan
area. For each participating individual, we use accelerometry data to
infer indicators of physical activity and GPS location data to
identify the region of residence. In this experimental setup, the
proposed approach achieves $81\%$ leave-one-out accuracy, with high
precision and recall for both classes.

Then, we demonstrate an application of the models in visualizing
heatmaps of the predicted population behavior for areas of
interest. The examination of the example heatmaps, in conjunction with
the environment characteristics, reveals useful and actionable
insights. The predictive models can also be directly adopted in urban
or policy design, especially for simulating imminent changes in the
population behavior as the outcome of possible policies and
interventions.

The rest of the paper is organized as follows. First, Section
\ref{sec:proposed_approach} presents an overview of the proposed
approach. Then, Section \ref{sec:experiments} demonstrates an
experimental application for predicting the physical activity level
for children and adolescent population. Section \ref{sec:heatmaps}
uses the predictive models to generate heatmaps of the
expected population behavior and provides a discussion using the
visualizations. Finally, Section \ref{sec:conclusions} concludes the
paper.

%%%%%%%%%%%%%%%%%%%%%%%%%%%%%%%%%%%%%%%%%%%%%%%%%%%%%%%%%%%%%%%%%%%%%%%%%%%%%%%%
\section{PROPOSED APPROACH}
\label{sec:proposed_approach}

\begin{figure}[!t]
\centering
  \includegraphics[width=0.8\linewidth]{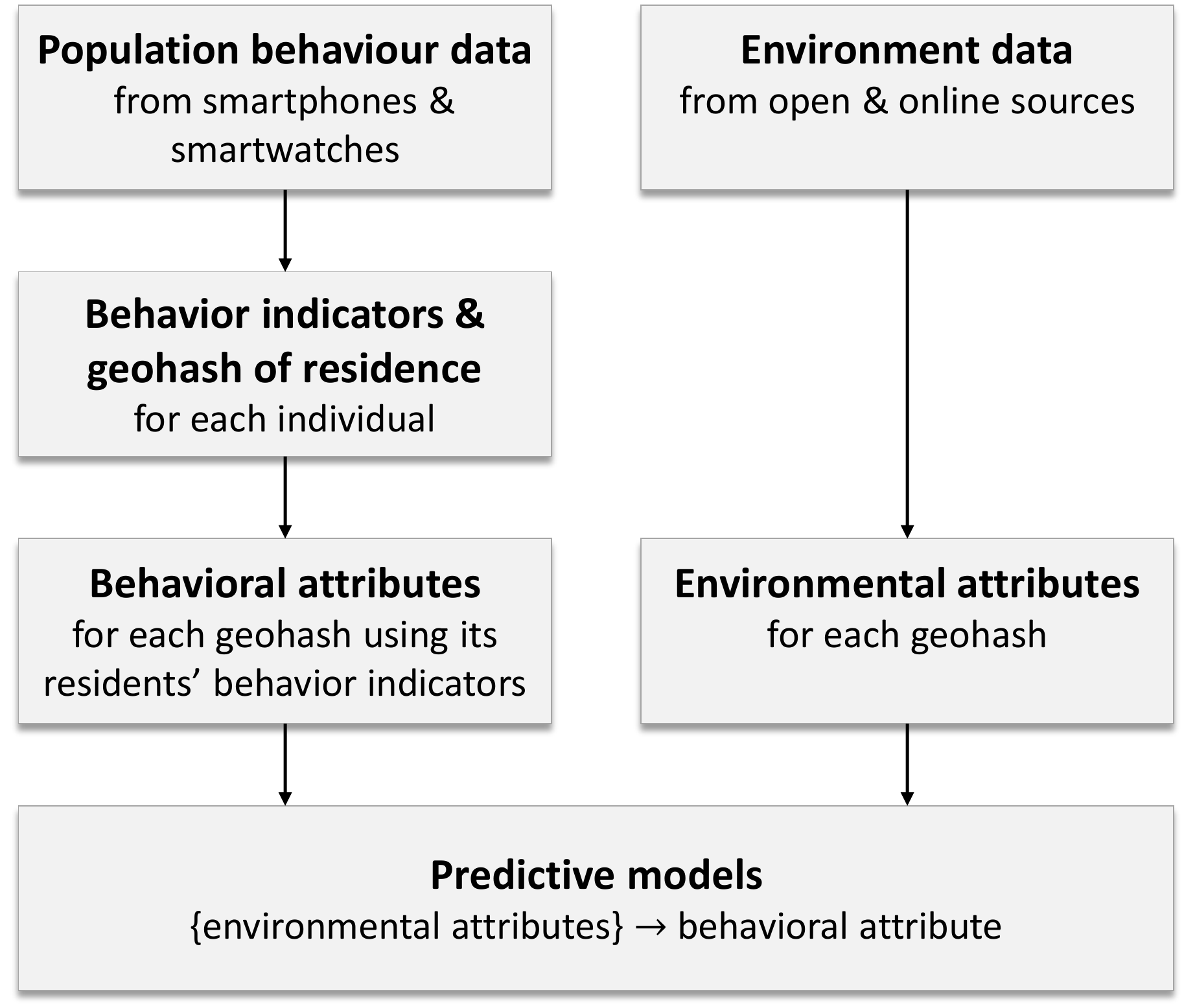}
  \vskip -5pt
  \caption{Schematic illustration of the proposed approach}
  \label{figure:approach}
\end{figure}

Figure \ref{figure:approach} sketches the proposed approach. The goal
is to build predictive models that map local attributes of the
environment to behaviors exhibited by the resident population.
This is accomplished through the following steps:

\begin{enumerate}
\item \emph{Large-scale behavioral data collection in real-life
  conditions} from the target population. We rely on commonly
  available sensors in smartphones and smartwatches, such as GPS and
  triaxial accelerometer. All behavioral data are recorded in
  real life, with no supervision. A smartphone application is used to
  collect and transmit the raw sensory data or calculated products.

\item \emph{Environment data collection} from external data
  sources. For example, data from GIS, such as the location and type
  for Points of Interests (POIs), the availability of open spaces,
  etc., or data from statistical authorities (e.g. unemployment rate,
  income).

\item For each participating individual, we extract \emph{behavioral
  indicators} from the collected sensory data. In this context, a
  behavioral indicator is an objectively measurable quantity for the
  behavior of one individual. For example, the average daily activity
  counts/min, the total steps, the types of the visited POIs, etc.

\item For each participating individual, we also identify the area of
  residence. For privacy protection, the residence location is
  assigned to a \emph{geohash} \cite{morton1966computer}. Geohash is a
  geocode system where geographic locations are mapped to a grid of
  rectangular areas. An advantage of this encoding is that we can
  easily tune the size of the rectangles by changing the hash length
  (shorter hashes outline larger areas); thus, enforcing the desired
  level of privacy protection.

\item We calculate \emph{behavioral attributes} for each geohash,
  which are aggregations for the behavioral indicators of the
  individuals living in the geohash. For example, the average of the
  activity counts/min indicator, the average daily steps, or the
  average number of weekly visits to fast food restaurants.

\item We calculate \emph{environmental attributes} for each
  geohash (also termed ``Local Extrinsic Conditions (LECs)'' in BigO
  system \cite{diou2019bigo_short}), using the available environment
  data. For example, the number of athletics facilities in the
  geohash, the number of fast food restaurants, or the local
  unemployment rate, etc.

\item Finally, we build predictive models that \emph{map environmental
  attributes to one or more behavioral attributes}. The premise
  is that there exist known, but unquantifiable, causal relationships
  between the average (expected) population behavior and the local
  environment.

\end{enumerate}

%%%%%%%%%%%%%%%%%%%%%%%%%%%%%%%%%%%%%%%%%%%%%%%%%%%%%%%%%%%%%%%%%%%%%%%%%%%%%%%%
\section{EXPERIMENTAL APPLICATION}
\label{sec:experiments}

This section demonstrates an experimental application of the proposed
approach. The models predict the behavioral attribute ``expected
physical activity level'' for relatively small geographic areas of
residence. The predicted classes are in High or Low level. A set of
urban environmental attributes is used as input to the models.

\subsection{Data and experimental setup}
\label{subsec:data_and_setup}
Table \ref{table:data} summarizes the data used in the
experiments. The data were collected from $410$ children and
adolescents (ages $9$-$18$) in the Thessaloniki metropolitan region,
Greece. The participants contributed data using the ``myBigO'' app
\cite{maramis2019developing_short} in the context of school
activities, under real-life conditions and without supervision during
the data collection period. The ``myBigO'' app collects self-reported
and sensory data (accelerometry and GPS location) from the smartphone
and, if available, a paired smartwatch
\cite{maramis2019developing_short}.  Typically, the data collection
period is $1$-$2$ weeks for each participating individual.

We applied two criteria to select the individuals for
the experiments. First, we selected the ones for which we were able to
clearly identify the geohash of the residence using
heuristics.\footnote{Based on GPS data recorded in the hours 23:00 to
  07:00.}  Second, we selected only individuals with more than 20
hours of active recording sessions spanning across at least 3
different days. As an active recording session we consider when the
individual is carrying the smartphone or wears a paired smartwatch
and, thus, extraction of physical activity indicators is possible.

\begin{table}[t]
  \caption{Overview of experiment data} 
  \renewcommand{\arraystretch}{1.1}
  \label{table:data}
  \vskip -5pt
  \begin{tabularx}{\linewidth}{Xr}
    \toprule
    {Total number of children} & $410$ \\
    {Total hours of recordings} & $90890$ \\
    {Number of children passing selection criteria} & $156$ \\
    {Male / Female / Other - Not specified} & $85 / 70 / 1 $ \\
    {Number of unique geohashes of residence (length 6)} & $42$ \\
    \bottomrule
  \end{tabularx}
\end{table}

From the $410$ participants in the Thessaloniki metropolitan region,
$156$ satisfied the selection criteria. The locations of residence
were found in $42$ different geohashes, using a 6 characters
length.\footnote{Approximately $610\text{m} \times 930\text{m}$ for
  Thessaloniki.}  For each individual, we extracted the target
behavioral indicator. In this experimental application, we calculated
the average activity counts per minute \cite{tryon1996fully} as the
indicator of physical activity, using the collected accelerometry data
of the active recording sessions.

\begin{figure}[!t]
\centering
  \includegraphics[width=0.85\linewidth]{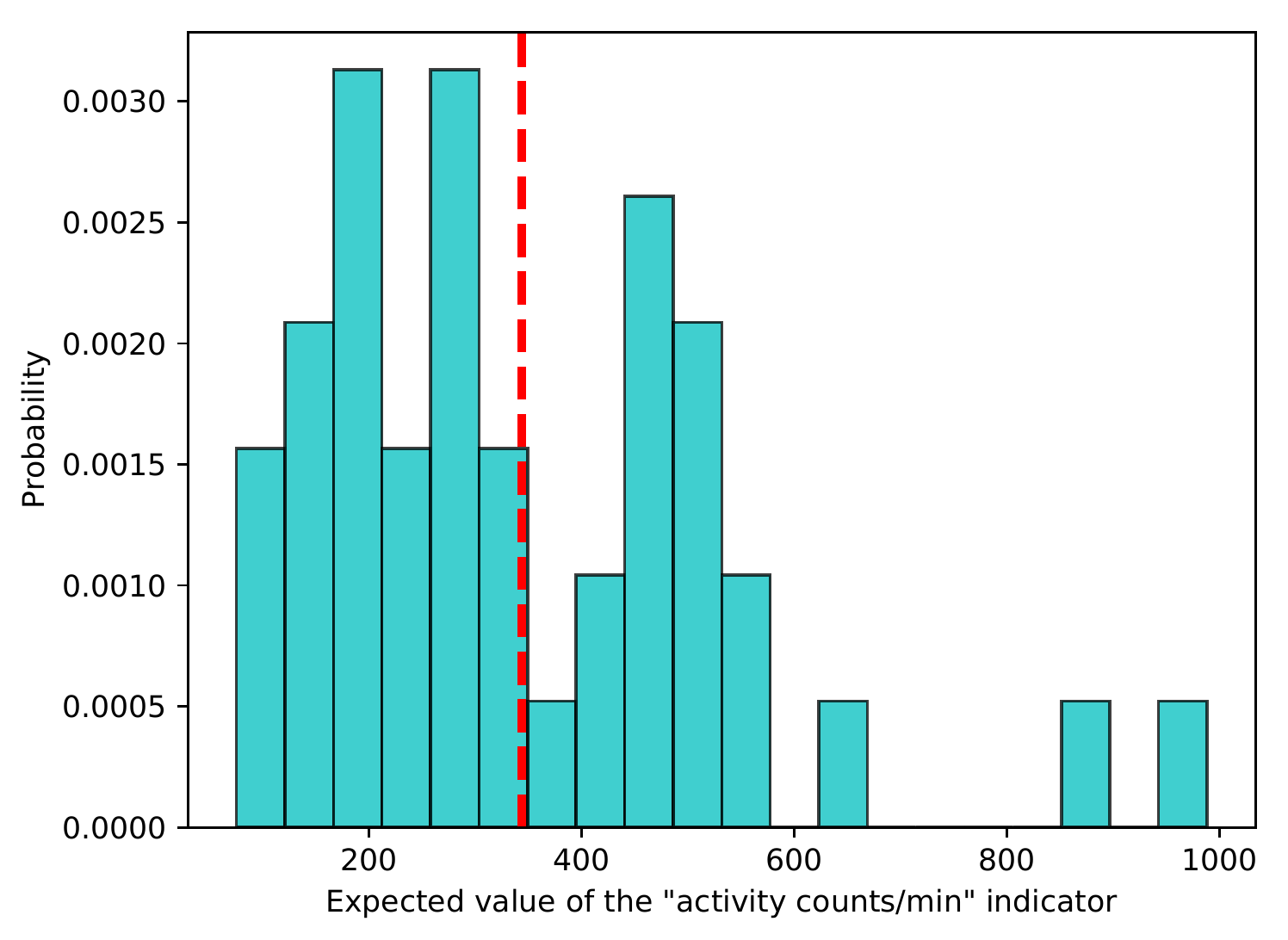}
  \vskip -5pt
  \caption{Histogram of the behavioral attribute for the
    geohashes. The dashed line indicates the distribution mean
    (approx. $343.9$ activity counts/min)}
  \label{figure:histogram}
\end{figure}

Then, for each geohash with residents, we averaged its residents'
behavioral indicators to calculate the behavioral attribute.  Thus,
the behavioral attribute is the expectation of the population
behavior in each geohash. The probability distribution of the
behavioral attribute is shown in Figure \ref{figure:histogram}.

Finally, for each geohash, we calculated a set of environmental
attributes from characteristics of the local urban
environment. Specifically, each environmental attribute corresponds to
the number of a specific POI type that is available in the geohash.
We calculated $4$ environmental attributes: ``athletics and sports,''
``fast food and take away restaurants,'' ``public parks'' and
``caf\'es and caf\'e bars.'' This information is provided from a
variety of online data sources, such as OpenStreetMap, Google Maps,
Foursquare, and Bing Maps.

\subsection{Model training}

We transformed the data to create a data set suitable for training
binary classification algorithms. For this purpose, the $i$-th
geohash, $i=1 \dots 42$, was converted into the data point $(\x_i,
y_i)$ that consists of:

\begin{itemize}
  \item A $4$-dimensional feature vector $\x_i$ that contains the
    environmental attributes. Each dimension is separately normalized
    in the range $[0, 1]$.
  \item A binary class label $y_i$, where: $y_i = \text{High}$ if
    behavioral attribute is greater than the distribution mean; $y_i =
    \text{Low}$ otherwise. The mean value of the distribution is
    approximately $343.9$ activity counts/min (see Figure
    \ref{figure:histogram}).
\end{itemize}

The predictive models are based on the random forests (RF) algorithm
\cite{breiman2001random}.  Through a $10$-fold cross-validation, we
selected the optimal hyperparameters for the number of decision trees,
the maximum number of levels in each tree and the measurement
criterion for the quality of split.

\subsection{Results}
\label{subsec:results}

Using the RF algorithm and the optimal hyperparameters, we applied a
leave-one-out (LOO) procedure to evaluate the prediction
performance. During LOO, one geohash (from the $42$ ones with
residents) was left out, an RF model was trained using the rest of the
geohashes, and the model's prediction was evaluated on the left-out
geohash.

Table \ref{table:confmat} shows the confusion matrix produced from the
LOO procedure and Table \ref{table:metrics} shows the corresponding
classification metrics.  The predictive models achieved high
leave-one-out accuracy of $81\%$. Furthermore, high precision and
recall was demonstrated for both classes.

Overall, the results verify that it is feasible to build accurate
generalizations for the population behavior in areas where only
environmental data are available.

\begin{table}[!t]
\centering
\caption{Confusion matrix for the LOO procedure. $\bf \text{Accuracy} = 81\%$}
\label{table:confmat}
\renewcommand{\arraystretch}{1.1}
\vskip -8pt
\begin{tabular}{c c c c }
&  & \multicolumn{2}{c}{\bf Predicted} \\ 
 & \multicolumn{1}{c|}{}  & Low & High \\
\cline{2-4}
\multirow{3}{*}{\rot{\bf Actual}} 
& \multicolumn{1}{c|}{Low}& $21$ & $3$ \\
& \multicolumn{1}{c|}{High}& $5$ & $13$ \\
\end{tabular}
\end{table}

\begin{table}[!t]
  \centering
  \caption{Classification metrics}
  \label{table:metrics}
  \renewcommand{\arraystretch}{1.1}
  \vskip -5pt
  \begin{tabular}{lllll}
    \toprule
    Class & & Precision & Recall & F1-score \\
    \midrule
    Low  & & $0.81$ & $0.88$ & $0.84$ \\
    High & & $0.81$ & $0.72$ & $0.76$ \\
    \bottomrule
  \end{tabular}
\end{table}

\begin{figure}[!b]
\centering
  \includegraphics[width=0.7\linewidth]{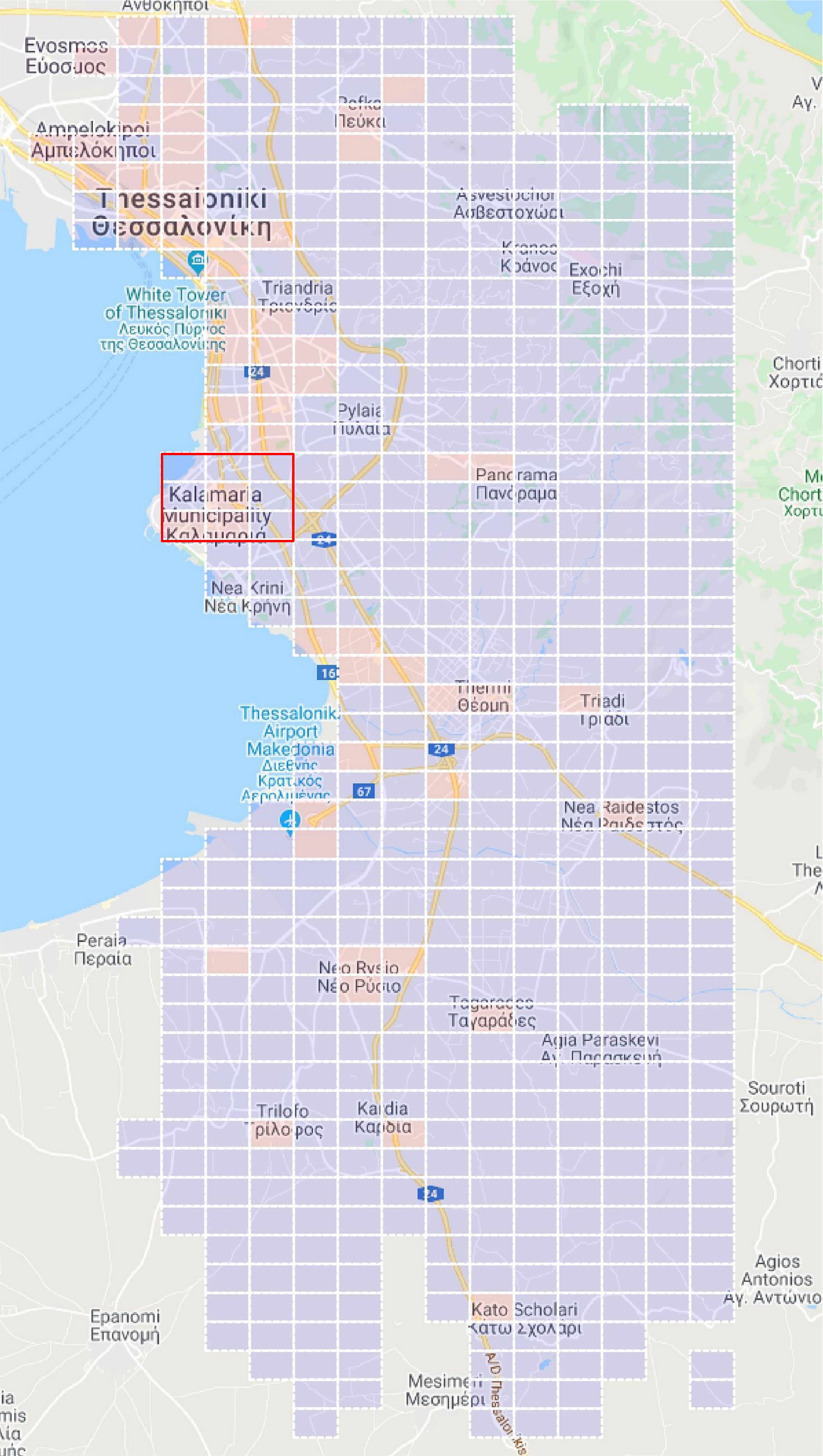} %BigO_Vis_Large_cropped_2
  \vskip -5pt
  \caption{Example heatmap for an area in the Thessaloniki
    metropolitan area. Grid is made from geohashes of hash length $6$.
    Geohashes predicted in the High class are colored red; geohashes
    predicted in the Low class are colored blue. The area marked with
    a red square is shown in detail in Figure
    \ref{figure:heatmap1}. Map data: Google.}
  \label{figure:heatmap_big}
\end{figure}

%%%%%%%%%%%%%%%%%%%%%%%%%%%%%%%%%%%%%%%%%%%%%%%%%%%%%%%%%%%%%%%%%%%%%%%%%%%%%%%%
\section{HEATMAP VISUALIZATIONS}
\label{sec:heatmaps}

\begin{figure*}[!th]
\centering
  \includegraphics[width=0.75\linewidth]{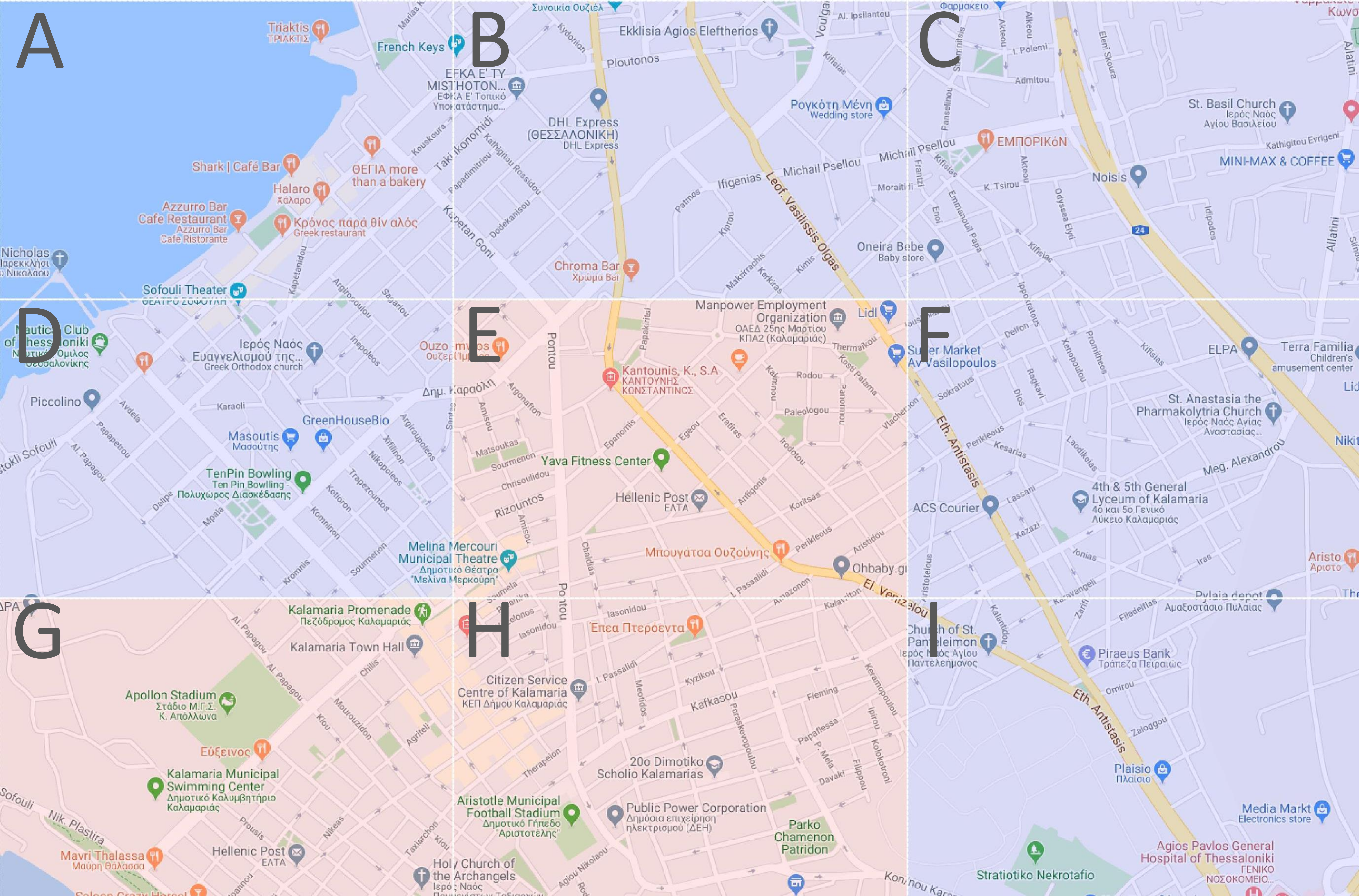}
  \vskip -5pt
  \caption{Example heatmap for a municipality region in Thessaloniki
    city. The figure shows $9$ geohashes (hash length $6$). Geohashes
    predicted in the High class are colored red; geohashes predicted
    in the Low class are colored blue. Map data: Google.}
  \label{figure:heatmap1}
\end{figure*}

In this section we exploit the high accuracy of the predictive models
to generate heatmaps of the expected population behavior in target
areas of interest.

Figure \ref{figure:heatmap_big} shows the model predictions per
geohash for a large part of the Thessaloniki metropolitan area. It is
worth noting that the original data span over $42$ geohashes, which is
a relatively small percentage of the total area of interest in the
Thessaloniki metropolitan region. Thus, using an imputation mechanism
--- such as the proposed predictive models --- is the only way we can
visualize complete and usable heatmaps.

Figure \ref{figure:heatmap1} shows a more targeted heatmap that covers
an area in a municipality. The geohashes E, G and H are predicted in the
High level class, which is colored red in the heatmap. In these
geohashes, there is a significant number of athletics/sports
facilities and public parks, providing exercise and active recreation
opportunities for the children and adolescents. Also, these areas have
an increased number of food outlets and caf\'es, which is an
indication of commercial and/or popular areas with pedestrian streets
and squares where people walk.

On the other hand, such urban environment characteristics are not
prominent in the other geohashes (A, B, C, D, F and I). These
geohashes are predicted in the Low level class, colored blue in
the heatmap. For example, geohash F is a fairly densely populated
area, however, it lacks POIs that would contribute to an active
lifestyle. Thus, from a public health authority's point-of-view, such
heatmaps provide direct indications for areas in need of attention.

%%%%%%%%%%%%%%%%%%%%%%%%%%%%%%%%%%%%%%%%%%%%%%%%%%%%%%%%%%%%%%%%%%%%%%%%%%%%%%%%
%% \section{CONCLUSIONS AND FUTURE WORK}
\section{CONCLUSIONS}
\label{sec:conclusions}

We presented a novel approach for building models that predict the
expected population behavior from the local environment in the regions
of residence. The proposed approach was demonstrated and evaluated in
the application of predicting the expected physical activity level for
children and adolescents population, using characteristics of the
urban environment as input. The experiment data were collected in the
context of BigO pilots from $156$ children and adolescents in the
Thessaloniki metropolitan region, Greece. The results verified the
feasibility of the proposed approach. In addition, we demonstrated an
application of the predictive models for visualizing heatmaps of the
population behavior, from which we derived useful insights. The
proposed approach can support further applications in urban design and
policy making, for example in simulating the effects of policies and
interventions on the population; however, further validation is needed
in these research directions.

\bibliographystyle{IEEEtran}

\begin{thebibliography}{10}
\providecommand{\url}[1]{#1}
\csname url@rmstyle\endcsname
\providecommand{\newblock}{\relax}
\providecommand{\bibinfo}[2]{#2}
\providecommand\BIBentrySTDinterwordspacing{\spaceskip=0pt\relax}
\providecommand\BIBentryALTinterwordstretchfactor{4}
\providecommand\BIBentryALTinterwordspacing{\spaceskip=\fontdimen2\font plus
\BIBentryALTinterwordstretchfactor\fontdimen3\font minus
  \fontdimen4\font\relax}
\providecommand\BIBforeignlanguage[2]{{%
\expandafter\ifx\csname l@#1\endcsname\relax
\typeout{** WARNING: IEEEtran.bst: No hyphenation pattern has been}%
\typeout{** loaded for the language `#1'. Using the pattern for}%
\typeout{** the default language instead.}%
\else
\language=\csname l@#1\endcsname
\fi
#2}}

\bibitem{who2016obesity}
{World Health Organisation}, ``Obesity and overweight, fact sheet,'' 2016,
  \url{https://www.who.int/news-room/fact-sheets/detail/obesity-and-overweight},
  Last accessed on 2020-01-22.

\bibitem{butte2007energy}
N.~F. Butte, E.~Christiansen, and T.~I.~A. Sørensen, ``Energy imbalance
  underlying the development of childhood obesity,'' \emph{Obesity}, vol.~15,
  no.~12, pp. 3056--3066, 2007.

\bibitem{romieu2017energy_short}
I.~Romieu \emph{et~al.}, ``Energy balance and obesity: what are the main
  drivers?'' \emph{Cancer Causes {\&} Control}, vol.~28, no.~3, pp. 247--258,
  2017.

\bibitem{world2019global}
{World Health Organization}, \emph{Global action plan on physical activity
  2018-2030: more active people for a healthier world}, 2019.

\bibitem{gilescorti2016city_short}
B.~Giles-Corti \emph{et~al.}, ``City planning and population health: a global
  challenge,'' \emph{The Lancet}, vol. 388, no. 10062, pp. 2912 -- 2924, 2016.

\bibitem{diou2019bigo_short}
C.~Diou \emph{et~al.}, ``A methodology for obtaining objective measurements of
  population obesogenic behaviors in relation to the environment,''
  \emph{Statistical Journal of the IAOS}, vol.~35, no.~4, pp. 677 -- 690, 2019.

\bibitem{sarafis2019behaviour}
I.~{Sarafis}, C.~{Diou}, and A.~{Delopoulos}, ``Behaviour profiles for
  evidence-based policies against obesity,'' in \emph{2019 41st Annual
  International Conference of the IEEE Engineering in Medicine and Biology
  Society (EMBC)}, July 2019, pp. 3596--3599.

\bibitem{morton1966computer}
G.~M. Morton, ``A computer oriented geodetic data base and a new technique in
  file sequencing,'' \emph{IBM Research}, 1966.

\bibitem{maramis2019developing_short}
C.~Maramis \emph{et~al.}, ``Developing a novel citizen-scientist smartphone app
  for collecting behavioral and affective data from children population,'' in
  \emph{8th EAI International Conference on Wireless Mobile Communication and
  Healthcare}, November 2019.

\bibitem{tryon1996fully}
W.~W. Tryon and R.~Williams, ``Fully proportional actigraphy: A new
  instrument,'' \emph{Behavior Research Methods, Instruments, {\&} Computers},
  vol.~28, no.~3, pp. 392--403, Sep 1996.

\bibitem{breiman2001random}
L.~Breiman, ``Random forests,'' \emph{Machine Learning}, vol.~45, no.~1, pp.
  5--32, Oct 2001.

\end{thebibliography}

\end{document}